\documentclass[10pt,twocolumn,english,aps,prd,superscriptaddress,nofootinbib,preprintnumbers,floatfix]{revtex4-2}

\usepackage[utf8]{inputenc}
\usepackage[T1]{fontenc}
\usepackage{lmodern}
\usepackage[english]{babel}
\usepackage{amsmath,amssymb,amsfonts,bm,mathrsfs}
\usepackage{graphicx}
\usepackage{xcolor}
\colorlet{BLUE}{blue} 
\usepackage{hyperref}
\usepackage{tikz}

\hypersetup{colorlinks=true,linkcolor=blue,citecolor=blue,urlcolor=blue}

\definecolor{lime}{HTML}{A6CE39}

\newcommand{\dd}{\mathrm{d}}
\newcommand{\ee}{\mathrm{e}}
\newcommand{\cO}{\mathcal{O}}
\newcommand{\Rar}{\mathcal{R}}
\newcommand{\bc}{b_{\rm c}}
\newcommand{\Om}{\Omega_{\rm c}}
\newcommand{\lam}{\lambda_{\rm c}}
\newcommand{\gamL}{\gamma_{\rm L}}

\begin{document}

\title{Critical curves and throat degeneracy in black bounces}

\author{Francisco S. N. Lobo}
\email{fslobo@ciencias.ulisboa.pt}
\affiliation{Instituto de Astrof\'{i}sica e Ci\^{e}ncias do Espa\c{c}o, Faculdade de Ci\^{e}ncias da Universidade de Lisboa, Edif\'{i}cio C8, Campo Grande, P-1749-016 Lisbon, Portugal}
\affiliation{Departamento de F\'isica, Faculdade de Ci\^encias da Universidade de Lisboa, Edif\'icio C8, Campo Grande, P-1749-016 Lisbon, Portugal}

\author{Diogo Mesquita}
\affiliation{Department of Physics, University of Coimbra, 3004-516 Coimbra, Portugal}

\author{Lu\'{i}s F. Dias da Silva} 
\email{fc53497@alunos.fc.ul.pt}
\affiliation{Instituto de Astrof\'{i}sica e Ci\^{e}ncias do Espa\c{c}o, Faculdade de Ci\^{e}ncias da Universidade de Lisboa, Edif\'{i}cio C8, Campo Grande, P-1749-016 Lisbon, Portugal}
\affiliation{Departamento de F\'isica, Faculdade de Ci\^encias da Universidade de Lisboa, Edif\'icio C8, Campo Grande, P-1749-016 Lisbon, Portugal}

\begin{abstract}
Identical critical curves need not imply identical null-orbit dynamics.
For a class of static spherical areal-radius deformations, we show
that a surviving original photon sphere retains its critical impact
parameter and orbital frequency, while its coordinate-time Lyapunov
exponent is multiplied by the radial Jacobian. The logarithmic
strong-deflection coefficient acquires the inverse factor. At a
reflection-symmetric throat, stability factorizes into a derivative of the original optical potential and the throat curvature. More generally, an order-$m$
extremum of the original potential composed with an order-$q$ areal-radius minimum produces
an optical degeneracy of order $p=mq$, with critical approach
$|x|\propto|t|^{-2/(p-2)}$ and reflected strong deflection
$\Delta\phi\propto(b/b_{\rm c}-1)^{-(p-2)/(2p)}$ for $p>2$.
For the Simpson--Visser family with mass $M$ and bounce parameter
$a$, the critical impact parameter remains Schwarzschild-like for
$a<3M$, although the instability varies with $a$. The merger at
$a=3M$ is quartic, separating stable and unstable throat branches.
These results isolate the null-orbit information preserved by a
radial deformation from that encoded in its local geometry.
\end{abstract}

\maketitle

\section{Introduction}
\label{sec:intro}

Horizon-scale images offer a direct way to study strong gravitational
lensing, but extracting geometry from an image requires separating the
null trajectories from the emitting matter. The Event Horizon
Telescope observations of M87* and Sgr~A* reveal compact ring-like
emission surrounding a central brightness depression
\cite{EHTM87I,EHTSgrAI}. The geometric critical curve organizes
strongly lensed trajectories; its existence and size do not alone
identify an event horizon or determine the surrounding intensity
pattern \cite{ClaudelVirbhadraEllis2001,PerlickTsupko2022}.
Different spacetimes can therefore share the same critical-curve scale
\cite{LimaCrispinoCunhaHerdeiro2021,BroderickSalehiGeorgiev2023}.
The relevant question is which additional null-orbit quantities can
retain information that this common scale conceals.

For a nondegenerate unstable circular null orbit, the critical impact
parameter measures the height of the optical potential, whereas the
Lyapunov exponent measures the local rate of radial separation.
Together with the orbital frequency, it controls the asymptotic
photon-ring hierarchy and the logarithmic strong-deflection law
\cite{Cardoso2009,Bozza2002,Johnson2020,GrallaLupsascaMarrone2020,
Kocherlakota2024,AratoreTsupkoPerlick2024}.
Their complementary metric dependence is established in spherical and
stationary settings
\cite{BroderickSalehiGeorgiev2023,SalehiBroderickGeorgiev2024,Salehi2025}.
The task here is to identify an exact deformation mechanism that
separates these quantities, rather than to establish this
complementarity for the first time.

Black-bounce spacetimes provide a useful setting because one parameter
can change the interior causal structure while leaving an exterior
critical scale unchanged. The Simpson--Visser geometry interpolates
between a regular black hole and a traversable wormhole
\cite{SimpsonVisser2019,Lobo2021}. The broader construction explored in Ref.~\cite{Lobo2021}
relates nonzero minima of the areal radius to regular black holes,
null transitions, and traversable wormholes, motivating a treatment
that separates the optical potential from the radial deformation.
Its shadows and disk images
\cite{LimaCrispinoCunhaHerdeiro2021,GuerreroOlmoRubieraSaez2021,Vagnozzi2023},
critical null orbits and characteristic frequencies
\cite{LimaJuniorBenoneCrispino2020,StuchlikVrba2021}, and
strong-deflection and throat-orbit structure
\cite{Tsukamoto2021,Tsukamoto2021Two,Tsukamoto2024}
have already been studied. In particular, the marginal throat at
$a=3M$ and its nonlogarithmic deflection are known
\cite{Tsukamoto2021,Tsukamoto2026,IgataSasakiTsukamoto2026,Tsukamoto2026Throat};
the metric-specific Lyapunov exponent has also appeared in an eikonal
analysis \cite{YangLan2026}. These results provide benchmarks for
a structural treatment of the underlying geometry.

The broader motivation extends beyond this example. Recent work
addresses bounds on shadow and instability observables
\cite{WaliaGeorgievChan2026}, imaging--ringdown relations
\cite{DiazGuerra2026}, and degeneracies lifted by higher-order optical
data \cite{LiuLiaoZhang2026,HanLiuLi2026}. Related black-bounce
studies examine optical topology \cite{MartinezGuerrero2026},
charged-black-bounce critical scales \cite{GuoMiao2022}, generalized
families \cite{Nascimento2026,Furtado2026}, and multithroat
configurations with equal impact scales but different instability
rates \cite{Pradhan2026}. Multiple optical and perturbative
barriers further emphasize the need to distinguish local orbit data
from global propagation \cite{SantosLessaMalufOlmo2026}.

We consider the restricted class in which the optical potential is
$U(x)=U_0[\Rar(x)]$, while the lapse and reciprocal radial metric
coefficient retain the same lapse function. This is a change of
geometry, not merely a radial coordinate transformation. Within this
class, we derive the exact inheritance of the critical impact
parameter and orbital frequency, the Jacobian rescaling of the
Lyapunov exponent, and a factorized throat-stability condition.
We also show that composing an order-$m$ extremum of $U_0(R)$ with an
order-$q$ areal-radius minimum produces an order-$mq$ optical
extremum. The resulting time-domain and strong-deflection exponents
organize the familiar quartic merger and its higher-order extensions.
Our contribution is this compact factorization and composition
framework; the individual Simpson--Visser optical features are
not claimed as new.

Throughout, stability means radial stability of a null geodesic.
It does not establish the stability of the spacetime or its supporting
matter. Stable light rings can support long-lived trapping and have
been associated with possible nonlinear instabilities in other
ultracompact-object settings
\cite{CardosoCrispinoMacedoOkawaPani2014,CunhaBertiHerdeiro2017};
applying such conclusions to a particular black bounce requires its
field equations and perturbative boundary conditions.
A separate example is the dynamical stability of black-bounce
thin-shell wormholes studied by Lobo, Simpson, and Visser
\cite{LoboSimpsonVisser2020}: there stability concerns radial
motion of a material shell and depends on its constitutive properties,
whereas the present analysis concerns null trajectories in a fixed
smooth geometry. Likewise, a
local critical impact parameter defines an observable critical curve
only when the corresponding orbit is accessible from the selected
asymptotic region. An observed darkness boundary additionally depends
on the illumination and absorption prescription.

We assume sufficiently smooth metric functions, circular orbits in
static regions with $F>0$, and a common asymptotic normalization of
the Killing time, with $F\to1$. For two asymptotically flat ends,
the areal radius satisfies
\begin{equation}
\Rar(x)=|x|+o(|x|),
\qquad
\Rar'(x)\to\pm1,
\qquad |x|\to\infty,
\label{eq:asymptoticR}
\end{equation}
with $\Rar(x)=x+o(x)$ and $\Rar'(x)\to1$ for a single end as
$x\to+\infty$. These conditions give the standard interpretation
of $b=|L|/E$ as the impact parameter at infinity. We use
$G=c=1$ and signature $(-+++)$.

The paper is organized as follows. Section~\ref{sec:general} defines
the deformation class and null dynamics. Section~\ref{sec:inheritance}
derives the inheritance relations and their strong-deflection
consequences. Sections~\ref{sec:throatgeneral} and
\ref{sec:criticalgeneral} establish throat stability and degeneracy
composition. Section~\ref{sec:SV} applies the results to the
Simpson--Visser family, and Sec.~\ref{sec:diagnostics} derives the
near-critical scalings and a model-dependent parameter inversion.
Section~\ref{sec:conclusion} summarizes the results and their
connections to future optical and perturbative studies.

\section{Areal-radius deformations and optical potential}
\label{sec:general}

Consider the static, spherically symmetric line element
\begin{equation}
\dd s^2=-F[\Rar(x)]\,\dd t^2+
\frac{\dd x^2}{F[\Rar(x)]}+\Rar^2(x)\,\dd\Omega^2,
\label{eq:generalmetric}
\end{equation}
where \(\dd\Omega^2=\dd\theta^2+\sin^2\theta\,\dd\phi^2\) and \(\Rar(x)>0\) is the areal radius. Here \(x\) is the radial coordinate of the deformed geometry, \(4\pi\Rar^2(x)\) is the invariant area of each symmetry two-sphere, and \(F(R)\) denotes the lapse function of the original geometry as a function of its areal radius \(R\). Thus \(F[\Rar(x)]\) means the original lapse evaluated at the deformed areal radius. Equation~\eqref{eq:generalmetric} is not the most general static spherical metric. It is the specific deformation class in which the lapse and reciprocal radial coefficient retain the same lapse function \(F\), while the areal radius entering them is replaced by \(R\mapsto\Rar(x)\).

It is important that Eq.~\eqref{eq:generalmetric} is \emph{not} obtained from the original geometry by a radial coordinate transformation. If one merely set \(R=\Rar(x)\) in the original metric, the radial term would become
\begin{equation}
\frac{\dd R^2}{F(R)}=
\frac{[\Rar'(x)]^2}{F[\Rar(x)]}\,\dd x^2,
\label{eq:coordinatechange}
\end{equation}
whereas Eq.~\eqref{eq:generalmetric} has \(g_{xx}=F^{-1}[\Rar(x)]\). Except where \(|\Rar'|=1\), the two metrics are therefore genuinely different. The Jacobian that appears below is consequently a property of the radial deformation, not a coordinate artifact.

For equatorial null geodesics, the Killing integrals are
\begin{equation}
E=F\dot t,
\qquad
L=\Rar^2\dot\phi,
\label{eq:EL}
\end{equation}
where a dot denotes differentiation with respect to an affine parameter. The constants \(E=-p_t\) and \(L=p_\phi\) are, respectively, the conserved Killing energy and azimuthal angular momentum of the null geodesic. We restrict to future-directed rays with \(E>0\) and quote the positive magnitudes \(b=|L|/E\) and \(\Omega=|\dd\phi/\dd t|\); choosing \(L>0\) is then only an orientation convention. Their common affine normalization is irrelevant to these ratios. Defining \(b=|L|/E\), the null condition becomes
\begin{equation}
\dot x^2=E^2\left[1-b^2U(x)\right],
\qquad
U(x)=\frac{F[\Rar(x)]}{\Rar^2(x)}.
\label{eq:radial}
\end{equation}
The corresponding original geometry is
\begin{equation}
\dd s_0^2=-F(R)\,\dd t^2+
\frac{\dd R^2}{F(R)}+R^2\,\dd\Omega^2,
\label{eq:originalmetric}
\end{equation}
and we define the original optical potential
\begin{equation}
U_0(R)=\frac{F(R)}{R^2},
\qquad
U(x)=U_0[\Rar(x)].
\label{eq:originalU}
\end{equation}
Primes on functions of \(x\) denote \(\dd/\dd x\), while subscripts \(R\) on quantities defined in the original geometry, such as \(U_{0,R}\) and \(U_{0,RR}\), denote derivatives with respect to the original areal radius \(R\).
A circular null orbit in a static region satisfies
\begin{equation}
U'(x_{\rm c})=0,
\qquad
b_{\rm c}^2=\frac{1}{U(x_{\rm c})},
\label{eq:circularconditions}
\end{equation}
with instability corresponding to \(U''(x_{\rm c})<0\) and stability to \(U''(x_{\rm c})>0\).

The coordinate angular frequency follows directly from Eq.~\eqref{eq:EL}:
\begin{equation}
\Omega_{\rm c}
=\left|\left.\frac{\dd\phi}{\dd t}\right|_{x_{\rm c}}\right|
=\frac{\sqrt{F[\Rar(x_{\rm c})]}}{\Rar(x_{\rm c})}
=\frac{1}{b_{\rm c}}.
\label{eq:Omegaidentity}
\end{equation}
With the asymptotic normalization specified above, $b_{\rm c}$
is the conserved impact parameter associated with the orbit and
$\Omega_{\rm c}$ is its coordinate-time orbital frequency.
Identifying this local impact scale with a critical curve seen from
a chosen asymptotic end additionally requires accessibility. Along a connecting static
radial path, $U(x)<U(x_{\rm c})$ must hold before the ray reaches
the selected orbit. A higher barrier forbids access, while an
intervening equal-height maximum can itself trap the critical ray.
Thus Proposition~1 preserves local orbit data; global barrier ordering
must be checked separately.

\section{Inheritance of critical-curve data and instability}
\label{sec:inheritance}

Suppose the original optical potential has a nondegenerate unstable photon sphere at \(R=R_\gamma\),
\begin{equation}
F(R_\gamma)>0,
\qquad
U_{0,R}(R_\gamma)=0,
\qquad
U_{0,RR}(R_\gamma)<0,
\label{eq:originalph}
\end{equation}
and suppose that the deformed geometry contains a point \(x_\gamma\) such that
\begin{equation}
\Rar(x_\gamma)=R_\gamma,
\qquad
\Rar'(x_\gamma)\neq0.
\label{eq:survive}
\end{equation}

\paragraph{Proposition 1 (critical-curve inheritance and instability splitting).}
Under Eqs.~\eqref{eq:originalph} and \eqref{eq:survive}, \(x_\gamma\) is a nondegenerate unstable circular null orbit of the deformed geometry. Its critical impact parameter and coordinate angular frequency are inherited exactly from the original geometry,
\begin{align}
\bc&=\frac{R_\gamma}{\sqrt{F(R_\gamma)}},
\label{eq:bcgeneral}\\
\Om&=\frac{\sqrt{F(R_\gamma)}}{R_\gamma}=\frac{1}{\bc},
\label{eq:Omegageneral}
\end{align}
whereas its coordinate-time Lyapunov exponent is
\begin{equation}
\lam=\left|\Rar'(x_\gamma)\right|\lambda_0,
\label{eq:inheritance}
\end{equation}
with Lyapunov exponent in the original geometry
\begin{equation}
\lambda_0^2=-\frac{F(R_\gamma)R_\gamma^2}{2}
U_{0,RR}(R_\gamma).
\label{eq:lambdaoriginal}
\end{equation}
Here \(\lambda_0>0\) is the coordinate-time Lyapunov exponent of the original photon sphere at \(R=R_\gamma\); correspondingly, \(b_{{\rm c},0}=R_\gamma/\sqrt{F(R_\gamma)}\) and \(\Omega_0=\sqrt{F(R_\gamma)}/R_\gamma\) denote the original critical impact parameter and orbital frequency.

\paragraph{Proof.}
Since \(U(x)=U_0[\Rar(x)]\),
\begin{equation}
U'(x)=U_{0,R}\Rar',
\label{eq:chain1}
\end{equation}
so Eq.~\eqref{eq:originalph} implies \(U'(x_\gamma)=0\). Differentiating once more gives
\begin{equation}
U''=U_{0,RR}(\Rar')^2+U_{0,R}\Rar'',
\end{equation}
and hence
\begin{equation}
U''(x_\gamma)=
[\Rar'(x_\gamma)]^2U_{0,RR}(R_\gamma)<0.
\label{eq:Usecond}
\end{equation}
Thus the maximum of $U_0(R)$ remains a nondegenerate maximum whenever \(\Rar'(x_\gamma)\neq0\). Equations~\eqref{eq:bcgeneral} and \eqref{eq:Omegageneral} follow because the value of \(U\) at the orbit is exactly the value in the original geometry. Finally, differentiating Eq.~\eqref{eq:radial} with respect to the affine parameter and linearizing around \(x=x_\gamma+\delta x\) yields
\begin{equation}
\delta\ddot x=-\frac{L^2}{2}U''(x_\gamma)\delta x.
\label{eq:affinelyap}
\end{equation}
To make the conversion from affine to coordinate time explicit, define
\(F_\gamma\equiv F(R_\gamma)\) and
\(U_\gamma\equiv U(x_\gamma)=F_\gamma/R_\gamma^{2}\). On the circular orbit,
\begin{equation}
\dot t=\frac{E}{F_\gamma},
\qquad
L^2=\frac{E^2}{U_\gamma}
=E^2\frac{R_\gamma^2}{F_\gamma}.
\label{eq:orbitnormalization}
\end{equation}
Because \(\dot x=0\) on the background orbit, variations of \(\dot t\) enter the radial perturbation equation only beyond linear order. Equation~\eqref{eq:affinelyap} may therefore be converted using the constant background value of \(\dot t\), giving
\begin{equation}
\frac{\dd^2\delta x}{\dd t^2}=\lam^2\delta x,
\qquad
\lam^2=-\frac{F(R_\gamma)R_\gamma^2}{2}U''(x_\gamma).
\label{eq:lambdageneral}
\end{equation}
Combining Eqs.~\eqref{eq:Usecond}, \eqref{eq:lambdaoriginal}, and \eqref{eq:lambdageneral} proves Eq.~\eqref{eq:inheritance}. \(\square\)

Proposition~1 isolates the exact factorization characteristic of the class \eqref{eq:generalmetric}: the height of the inherited optical-potential maximum, and therefore \(b_{\rm c}\) and \(\Omega_{\rm c}\), is fixed by the original geometry, while the curvature of the maximum with respect to the deformation coordinate \(x\) acquires the Jacobian factor. This is the restricted-class mechanism underlying a broader complementarity between critical-curve and photon-ring observables established in more general spherical and stationary settings \cite{BroderickSalehiGeorgiev2023,SalehiBroderickGeorgiev2024}.

\subsection{Strong-deflection coefficient and photon-ring hierarchy}

For a specified nondegenerate unstable circular null orbit, the local logarithmic strong-deflection coefficient \(\bar a\) follows from the expansion of the orbit equation
\begin{equation}
\frac{\dd\phi}{\dd x}
=\frac{b}{\Rar^2(x)\sqrt{1-b^2U(x)}}.
\label{eq:orbitphi}
\end{equation}
We define \(\bar a>0\) as the coefficient of the local logarithmic divergence of the accumulated azimuthal angle on the scattering branch associated with this orbit,
\begin{equation}
\Delta\phi(b)=-\bar a\ln\!\left(\frac{b}{b_{\rm c}}-1\right)+\cO(1),
\qquad b\to b_{\rm c}^{+},
\label{eq:abarDefinition}
\end{equation}
where any convention-dependent finite subtraction in the deflection angle is contained in the \(\cO(1)\) term.
Here the scattering branch is a ray reflected back into the same
exterior, with its turning point approaching the selected maximum
and with one incoming and one outgoing near-orbit leg. The integral
at exactly $b=b_{\rm c}$ diverges; expanding the near-critical
integral and extracting its logarithmic coefficient gives
\begin{equation}
\bar a=\frac{1}{R_\gamma^2}
\sqrt{-\frac{2}{U''(x_\gamma)}},
\label{eq:abarlocal}
\end{equation}
and Eqs.~\eqref{eq:Omegageneral} and \eqref{eq:lambdageneral} give
\begin{equation}
\bar a=\frac{\Om}{\lam}.
\label{eq:abaromegalambda}
\end{equation}
The relation \(\bar a=\Omega_{\rm c}/\lambda_{\rm c}\) itself has precedent in the connection between strong-deflection lensing and eikonal quasinormal modes \cite{StefanovYazadjievGyulchev2010}, and its imaging--ringdown interpretation has recently been revisited for modified spherical black holes \cite{DiazGuerra2026}. The same relation may be interpreted in terms of local geometric invariants, emphasizing that \(\bar a\) contains information beyond the critical-curve scale \cite{Igata2026}. The new statement in the present restricted class is its exact transformation under an areal-radius deformation. Proposition~1 therefore gives
\begin{equation}
\bar a=\frac{\bar a_0}{|\Rar'(x_\gamma)|},
\label{eq:abarinherit}
\end{equation}
where \(\bar a_0=\Omega_0/\lambda_0\) is the coefficient for the original geometry.

Defining the half-orbit lensing exponent $\gamL=\pi \lam/\Om=\pi/\bar a$,
where \(\gamL\) is dimensionless and one increment of the high-order image index corresponds asymptotically to one additional half-orbit around the critical null orbit, the widths, separations, and fluxes of successive sufficiently high-order photon-ring images inherit exponential scaling controlled by \(\gamL\), schematically
\begin{equation}
\frac{w_{n+1}}{w_n}\sim\ee^{-\gamL},
\qquad n\gg1,
\label{eq:ringdemag}
\end{equation}
Here \(w_n\) denotes a representative width of the \(n\)th sufficiently high-order photon-ring subimage; analogous exponential factors govern other local hierarchy measures, up to source- and geometry-dependent prefactors \cite{Johnson2020,GrallaLupsascaMarrone2020,Kocherlakota2024,AratoreTsupkoPerlick2024,Salehi2025}. In horizonless or multibarrier geometries, Eq.~\eqref{eq:abaromegalambda} characterizes only the local logarithmic divergence associated with the chosen unstable orbit; finite strong-deflection terms, additional inner images, and global image branches depend on the full potential structure \cite{ShaikhBanerjeePaulSarkar2019,Tsukamoto2021Two,SantosLessaMalufOlmo2026}.

\section{Reflection-symmetric throats}
\label{sec:throatgeneral}

Consider a reflection-symmetric geometry for which
\begin{align}
\Rar(-x)&=\Rar(x),
&\Rar(0)&=R_0>0,\nonumber\\
\Rar'(0)&=0,
&\Rar''(0)&=\Rar_2>0.
\label{eq:throatassumptions}
\end{align}
The surface \(x=0\) is a nondegenerate minimum of the areal radius. Reflection symmetry makes the throat a stationary point of the composed optical potential, but it does not by itself determine whether the stationary null orbit is physical or stable.

\paragraph{Proposition 2 (factorized throat stability).}
If \(F(R_0)>0\), the throat lies in a static region and supports a circular null orbit with
\begin{equation}
 b_{\rm th}=\frac{R_0}{\sqrt{F(R_0)}},
\qquad
\Omega_{\rm th}=\frac{\sqrt{F(R_0)}}{R_0}.
\label{eq:throatbOmega}
\end{equation}
Its second derivative factorizes as
\begin{equation}
U''(0)=U_{0,R}(R_0)\Rar_2.
\label{eq:throatsecond}
\end{equation}
Hence the throat orbit is stable when \(U_{0,R}(R_0)>0\), unstable when \(U_{0,R}(R_0)<0\), and marginal when \(U_{0,R}(R_0)=0\).

For the nondegenerate cases, define
\begin{equation}
\Xi_0=-\frac{F(R_0)R_0^2\Rar_2}{2}U_{0,R}(R_0).
\label{eq:Xithroat}
\end{equation}
Then \(\Xi_0>0\) gives the coordinate-time unstable exponent
\begin{equation}
\lambda_{\rm th}=\sqrt{\Xi_0},
\label{eq:throatlambda}
\end{equation}
whereas \(\Xi_0<0\) gives the coordinate-time small-oscillation frequency $\omega_{\rm th}=\sqrt{-\Xi_0}$.

The general classification of circular light orbits at \(Z_2\)-symmetric wormhole throats is known \cite{Tsukamoto2024}. Proposition~2 is the specialization to the compositional class \eqref{eq:generalmetric}: all dependence on the deformation near a nondegenerate throat enters through the positive factor \(\Rar_2\), so the stability sign is inherited from the single derivative of the original potential \(U_{0,R}(R_0)\). The condition \(F(R_0)>0\) is essential. In a black-hole branch the minimum-area surface may lie in a nonstatic region, in which case it should not be interpreted as an ordinary stationary photon sphere even though \(U'(0)=0\) algebraically.

The terms ``stable'' and ``unstable'' in Proposition~2 refer exclusively to the radial null-geodesic effective potential. In particular, geodesic stability does not establish dynamical stability of the background. In related horizonless ultracompact-object settings, stable light rings can support long-lived trapped modes and have been associated with possible nonlinear instability \cite{CardosoCrispinoMacedoOkawaPani2014,CunhaBertiHerdeiro2017}; establishing either conclusion here would require the full perturbation problem.

\section{Degeneracy composition at a symmetric throat}
\label{sec:criticalgeneral}

The nondegenerate inheritance regime ends when a original photon
sphere is driven onto an extremum of the areal radius: its Lyapunov
exponent tends to zero, while the associated logarithmic coefficient
diverges. The compositional form \(U(x)=U_0[\Rar(x)]\) makes the resulting degeneracy especially transparent.

Let \(U_\gamma\equiv U_0(R_\gamma)>0\) denote the original optical-potential value at the critical areal radius. The integers \(m\) and \(q\) introduced below are the orders of the first nonvanishing local variations of the original optical potential and throat areal radius, respectively, while \(C_m\) and \(D_q\) are their corresponding leading coefficients.

Suppose the original optical potential first departs from its critical value at order \(m\ge2\),
\begin{equation}
U_0(R)-U_\gamma
=C_m(R-R_\gamma)^m
+o\!\left[(R-R_\gamma)^m\right],
\label{eq:generalPotentialDeg}
\end{equation}
with \(C_m\neq0\). For an unstable maximum of $U_0(R)$, \(m\) is even and \(C_m<0\). We also assume \(U_\gamma>0\), so that the critical impact parameter is real. Suppose further that a reflection-symmetric throat satisfies \(\Rar(0)=R_\gamma\) and first departs from its minimum at even order \(q\ge2\),
\begin{equation}
\Rar(x)-R_\gamma
=D_qx^q+o(x^q),
\qquad D_q>0.
\label{eq:generalThroatDeg}
\end{equation}

\paragraph{Proposition 3 (multiplicative degeneracy-composition theorem).}
Under Eqs.~\eqref{eq:generalPotentialDeg} and \eqref{eq:generalThroatDeg}, the composed optical potential satisfies
\begin{equation}
U(x)-U_\gamma
=C_mD_q^{\,m}x^{p}+o(x^p),
\qquad
p=mq.
\label{eq:multiplicativeDeg}
\end{equation}
Thus the order of the throat-centered optical-potential degeneracy is the product of the order of the original potential extremum and the areal-radius-minimum order. For a smooth local maximum of the original optical potential the first nonzero order \(m\) is even, while reflection symmetry makes \(q\) even. Hence, in the smooth symmetric setting considered here, \(p=mq\) is necessarily a multiple of four; the quartic case is the least-degenerate possibility.

\paragraph{Proof.}
Equation~\eqref{eq:multiplicativeDeg} follows directly by substituting Eq.~\eqref{eq:generalThroatDeg} into Eq.~\eqref{eq:generalPotentialDeg}. \(\square\)

When \(C_m<0\), the throat-centered orbit is a marginally unstable maximum of order \(p\). At the critical impact parameter \(b_{\rm c}^2=1/U_\gamma\), define
\begin{equation}
\kappa_p=F(R_\gamma)
\sqrt{-\frac{C_mD_q^{\,m}}{U_\gamma}}.
\label{eq:kappap}
\end{equation}
Then the critical radial motion obeys
\begin{equation}
\left|\frac{\dd x}{\dd t}\right|
\simeq \kappa_p |x|^{p/2},
\label{eq:generalCriticalMotion}
\end{equation}
and therefore
\begin{equation}
|x(t)|\sim
\left[\frac{p-2}{2}\,\kappa_p|t-t_0|\right]^{-2/(p-2)}.
\label{eq:generalTimeScaling}
\end{equation}
The constant \(t_0\) fixes the arbitrary origin of coordinate time along the asymptotic critical trajectory.
The exponential Lyapunov regime is recovered only for a nondegenerate quadratic maximum; for every \(p>2\), the critical approach is algebraic.

The same local normal form fixes the leading strong-deflection power. For a scattering branch with \(\delta=b/b_{\rm c}-1\to0^+\), the near-critical contribution to the orbit integral rescales as
\begin{equation}
\Delta\phi_{\rm div}
\propto \delta^{-(p-2)/(2p)},
\qquad p>2,
\label{eq:generalDeflectionScaling}
\end{equation}
with a branch- and geometry-dependent coefficient. This exponent follows by balancing \(x^p\) against \(\delta\) in the denominator of Eq.~\eqref{eq:orbitphi} and is consistent with the general strong-deflection theory of marginal and degenerate photon spheres \cite{Tsukamoto2026,IgataSasakiTsukamoto2026,Tsukamoto2026Throat,Tsukamoto2020}.


The generic collision considered in the original Simpson--Visser problem corresponds to a nondegenerate maximum of $U_0(R)$ and a nondegenerate symmetric throat minimum,
\begin{equation}
U_{0,R}(R_\gamma)=0,
\qquad
U_{0,RR}(R_\gamma)<0,
\qquad
\Rar''(0)=\Rar_2>0.
\label{eq:collisionassumptions}
\end{equation}
Near the throat,
\begin{equation}
\Rar(x)=R_\gamma+\frac12\Rar_2x^2+\cO(x^4),
\label{eq:Rexpansion}
\end{equation}
while
\begin{equation}
U_0(R)=U_\gamma+\frac12U_{0,RR}(R_\gamma)(R-R_\gamma)^2
+\cO[(R-R_\gamma)^3].
\label{eq:U0expansion}
\end{equation}
Thus \(m=q=2\), so Proposition~3 gives a quartic degeneracy,
\begin{equation}
U'(0)=U''(0)=U^{(3)}(0)=0,
\label{eq:triplezero}
\end{equation}
with
\begin{equation}
U^{(4)}(0)=3U_{0,RR}(R_\gamma)\Rar_2^2<0,
\label{eq:Ufourth}
\end{equation}
and
\begin{equation}
U(x)=U_\gamma+
\frac18U_{0,RR}(R_\gamma)\Rar_2^2x^4+\cO(x^6).
\label{eq:quarticgeneric}
\end{equation}
The ordinary Lyapunov exponent therefore vanishes even though the orbit remains marginally unstable. Equation~\eqref{eq:generalTimeScaling} reduces to
\begin{equation}
|x(t)|\simeq
\frac{2}{\Rar_2\lambda_0|t-t_0|},
\label{eq:algebraicgeneral}
\end{equation}
while Eq.~\eqref{eq:generalDeflectionScaling} gives the characteristic quartic strong-deflection scaling
\begin{equation}
\Delta\phi_{\rm div}
\propto \left|\frac{b}{b_{\rm c}}-1\right|^{-1/4}.
\label{eq:quarticDeflectionScaling}
\end{equation}
If either the extremum of $U_0(R)$ or the throat minimum is more highly degenerate, Proposition~3 gives the corresponding higher order directly.

\section{Simpson--Visser optical phase diagram}
\label{sec:SV}

The Simpson--Visser spacetime \cite{SimpsonVisser2019}, whose null-orbit, absorption, circular-geodesic, shadow, and lensing properties have been developed in Refs.~\cite{LimaCrispinoCunhaHerdeiro2021,GuerreroOlmoRubieraSaez2021,Tsukamoto2021,LimaJuniorBenoneCrispino2020,StuchlikVrba2021,MartinezGuerrero2026}, is
\begin{equation}
\dd s^2=-A(x)\dd t^2+\frac{\dd x^2}{A(x)}+R^2(x)\dd\Omega^2,
\label{eq:SVmetric}
\end{equation}
with
\begin{equation}
R(x)=\sqrt{x^2+a^2},
\qquad
A(x)=1-\frac{2M}{R(x)}.
\label{eq:SVfunctions}
\end{equation}
Here \(M>0\) is the asymptotic mass parameter (equal to the ADM mass at either asymptotically flat end), \(a\ge0\) is the bounce parameter and minimum areal radius \(R(0)=a\), \(A(x)\) is the static lapse function, and \(x=0\) is the bounce/throat surface.
For $a>0$, the global coordinate runs over $-\infty<x<+\infty$.
The case $a=0$ is the Schwarzschild limit, with $R=|x|$;
$x=0$ is then singular and is not a regular throat. In that limit
we use one exterior, for example $x>2M$. For $0<a<2M$ the geometry is a regular black hole with horizons
\begin{equation}
x_\pm=\pm\sqrt{4M^2-a^2},
\label{eq:SVhorizons}
\end{equation}
\(a=2M\) is the null-throat (one-way wormhole) black-bounce transition, and \(a>2M\) is a two-way traversable wormhole \cite{SimpsonVisser2019,Tsukamoto2024}.

\begin{figure}[tb]
\includegraphics[width=\columnwidth]{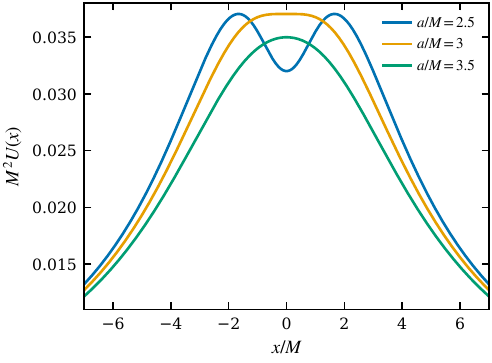}
\caption{Optical potential \(M^2U(x)\) for representative Simpson--Visser wormholes. For \(2M<a<3M\), two unstable off-throat photon spheres coexist with a stable throat circular null orbit. At \(a=3M\) the three stationary structures coalesce into a quartic marginally unstable orbit. For \(a>3M\) the throat is the unique unstable photon sphere.}
\label{fig:potential}
\end{figure}

The lapse function is that of Schwarzschild spacetime,
\begin{equation}
F(R)=1-\frac{2M}{R},
\end{equation}
and the optical potential is
\begin{equation}
U(x)=\frac{1}{R^2}\left(1-\frac{2M}{R}\right)
=\frac{R-2M}{R^3}.
\label{eq:SVU}
\end{equation}
Its derivative is
\begin{equation}
U'(x)=\frac{2x(3M-R)}{R^5}.
\label{eq:SVUprime}
\end{equation}
The derivative identities needed to evaluate the inherited and throat-centered branches are
\begin{align}
R'(x)&=\frac{x}{R},
&R''(x)&=\frac{a^2}{R^3},\nonumber\\
U_{0,R}(R)&=\frac{2(3M-R)}{R^4},
&U_{0,RR}(R)&=\frac{6(R-4M)}{R^5}.
\label{eq:SVderivatives}
\end{align}
In particular,
\begin{equation}
U_{0,RR}(3M)=-\frac{2}{81M^4},
\end{equation}
so Eq.~\eqref{eq:lambdaoriginal} gives the Schwarzschild value
\(\lambda_0=(3\sqrt3M)^{-1}\). Together with \(R'(x_\gamma)=x_\gamma/(3M)\), this supplies all intermediate ingredients for the off-throat result below.
For $a>0$, a stationary point always occurs algebraically at
$x=0$, although it is a stationary circular null orbit only when
the surface is static. For \(a<3M\), two additional off-throat extrema occur at
\begin{equation}
R_\gamma=3M,
\qquad
x_\gamma=\pm\sqrt{9M^2-a^2}.
\label{eq:SVph}
\end{equation}
At \(a=3M\) these two extrema coalesce with the throat. In the black-hole branch \(0<a<2M\), each asymptotic exterior has the corresponding unstable photon sphere at areal radius \(R=3M\); for a chosen exterior one takes the appropriate sign of \(x_\gamma\). The complete light-ring phase structure is summarized in Fig.~\ref{fig:potential} and Table~\ref{tab:phase}.

\subsection{Critical Curve-degenerate off-throat branch}
\label{subsec:offthroat}

For \(a<3M\), the off-throat photon spheres satisfy \(R_\gamma=3M\), so
\begin{equation}
U(x_\gamma)=\frac{1}{27M^2}.
\end{equation}
Proposition~1 therefore gives
\begin{equation}
\bc=3\sqrt3M,
\qquad
\Om=\frac{1}{3\sqrt3M}
.
\label{eq:SVshadowOmega}
\end{equation}
The critical curve is exactly Schwarzschild-like throughout the entire branch, including both the regular-black-hole regime \(0<a<2M\) and the traversable-wormhole interval \(2M<a<3M\). This metric-specific invariance of the critical impact parameter is known from earlier Simpson--Visser optical studies \cite{LimaCrispinoCunhaHerdeiro2021,LimaJuniorBenoneCrispino2020,StuchlikVrba2021}; here it is recovered as a direct corollary of Proposition~1.

The radial Jacobian at the orbit is
\begin{equation}
|R'(x_\gamma)|
=\frac{|x_\gamma|}{3M}
=\sqrt{1-\frac{a^2}{9M^2}},
\label{eq:SVjacobian}
\end{equation}
so
\begin{equation}
\lam=\frac{1}{3\sqrt3M}
\sqrt{1-\frac{a^2}{9M^2}}
.
\label{eq:lambdaSV}
\end{equation}
The corresponding strong-deflection and ring-hierarchy parameters are
\begin{align}
\bar a=\frac{1}{\sqrt{1-a^2/(9M^2)}}
,
\label{eq:abarSVsub}\\
\frac{\gamL}{\pi}=\frac{\lam}{\Om}
=\sqrt{1-\frac{a^2}{9M^2}}
.
\label{eq:gammaSVsub}
\end{align}
The metric-specific Lyapunov expression agrees with the recent eikonal analysis of Ref.~\cite{YangLan2026}; the structural result here is that it follows immediately from the general inheritance relation \eqref{eq:inheritance}.

\subsection{Stable throat orbit in the traversable interval}
\label{subsec:stablethroat}

At the throat \(R_0=a\), Eq.~\eqref{eq:SVderivatives} gives \(R''(0)=1/a\) and \(U_{0,R}(a)=2(3M-a)/a^4\). Substitution into Proposition~2 gives
\begin{equation}
U''(0)=\frac{2(3M-a)}{a^5}.
\label{eq:SVthroatsecond}
\end{equation}
The throat is static only for \(a>2M\). Hence the physical throat circular null orbit is stable precisely in $2M<a<3M$.
Its impact parameter and coordinate frequency are
\begin{equation}
 b_{\rm th}=\frac{a}{\sqrt{1-2M/a}},
\qquad
\Omega_{\rm th}=\frac{\sqrt{1-2M/a}}{a},
\label{eq:stablebOmega}
\end{equation}
and its small radial oscillation frequency is
\begin{equation}
\omega_{\rm th}=\frac{\sqrt{(a-2M)(3M-a)}}{a^2}
.
\label{eq:stablefreq}
\end{equation}
The throat orbit is distinct from the outer unstable separatrix. Indeed,
\begin{equation}
	b_{\rm th}^2-27M^2
	=
	\frac{(a-3M)^2(a+6M)}{a-2M}
	>0,
	\quad
	2M<a<3M.
	\label{eq:bthgap}
\end{equation}
So the stable throat orbit has a larger circular-orbit impact parameter and a smaller coordinate angular frequency than the off-throat unstable photon sphere.
This stable throat orbit is the antiphoton-sphere branch identified in the general classification of reflection-symmetric wormholes \cite{Tsukamoto2024}. For \(a\le2M\), the stationary point at \(x=0\) lies on or inside the nonstatic black-bounce region and should not be interpreted as an ordinary stable photon orbit.

\subsection{Quartic merger at \texorpdfstring{\(a=3M\)}{a=3M}}
\label{subsec:critical}

At \(a=3M\), the two off-throat unstable photon spheres and the stable throat orbit coalesce at \(x=0\). A direct expansion of Eq.~\eqref{eq:SVU} gives
\begin{equation}
U(x)=\frac{1}{27M^2}-\frac{x^4}{2916M^6}+\cO(x^6)
.
\label{eq:quarticSV}
\end{equation}
Thus
\begin{equation}
U''(0)=U'''(0)=0,
\qquad
U''''(0)=-\frac{2}{243M^6}<0.
\label{eq:SVfourth}
\end{equation}
The critical impact parameter remains \(3\sqrt3M\), but the ordinary Lyapunov exponent vanishes. At \(b=3\sqrt3M\),
\begin{equation}
\frac{\dd x}{\dd t}\simeq
\pm\frac{x^2}{18\sqrt3M^2},
\end{equation}
so
\begin{equation}
|x(t)|\simeq\frac{18\sqrt3M^2}{|t-t_0|}
.
\label{eq:SVquarticmotion}
\end{equation}
The critical orbit is therefore marginally unstable, with algebraic rather than exponential approach. Proposition~3 fixes the corresponding leading strong-deflection scaling to \(\Delta\phi_{\rm div}\propto|b/b_{\rm c}-1|^{-1/4}\), in agreement with the known nonlogarithmic Simpson--Visser behavior \cite{Tsukamoto2021,Tsukamoto2026Throat}.

The local bifurcation is seen directly by expanding the optical potential at fixed \(a\):
\begin{equation}
U(x;a)=\frac{a-2M}{a^3}
+\frac{3M-a}{a^5}x^2
+\frac{4a-15M}{4a^7}x^4
+\cO(x^6).
\label{eq:normalform}
\end{equation}
The quadratic coefficient changes sign at \(a=3M\), while the quartic coefficient is negative there. This is the local normal form of the merger of a stable central extremum with the two symmetric unstable extrema.

\subsection{Supercritical throat branch}
\label{subsec:supercritical}

For \(a>3M\), the off-throat roots disappear and the throat is the unique unstable photon sphere. From Eqs.~\eqref{eq:throatbOmega}, \eqref{eq:SVthroatsecond}, and \eqref{eq:throatlambda},
\begin{align}
\bc=\frac{a}{\sqrt{1-2M/a}}
,
\label{eq:bcSuper}\\
\Om=\frac{\sqrt{1-2M/a}}{a}
,
\label{eq:OmegaSuper}\\
\lam=\frac{\sqrt{(a-2M)(a-3M)}}{a^2}
.
\label{eq:lambdaSuper}
\end{align}
The last result follows explicitly from
\begin{equation}
\lambda_{\rm th}^2
=-\frac12\left(1-\frac{2M}{a}\right)a^2U''(0)
=\frac{(a-2M)(a-3M)}{a^4}.
\label{eq:lambdaSuperSquared}
\end{equation}
For \(2M<a<3M\), the same linearized equation changes from exponential growth to harmonic motion and yields Eq.~\eqref{eq:stablefreq}.
Consequently,
\begin{align}
\bar a=\frac{1}{\sqrt{1-3M/a}}
,
\label{eq:abarSuper}\\
\frac{\gamL}{\pi}=\sqrt{1-\frac{3M}{a}}
.
\label{eq:gammaSuper}
\end{align}
Unlike the branch \(a<3M\), the critical-curve scale now depends explicitly on the bounce parameter.

\begin{figure}[tb]
\includegraphics[width=\columnwidth]{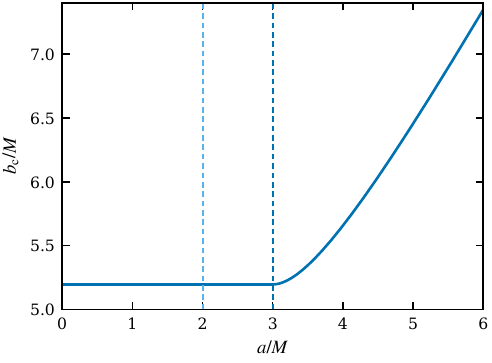}
\caption{Critical impact parameter in the Simpson--Visser family. The branch \(\bc/M=3\sqrt3\) is exactly critical-curve degenerate with Schwarzschild up to \(a=3M\). For \(a>3M\) the critical orbit is throat-centered and the critical-curve scale grows with \(a\). Vertical lines mark \(a=2M\) and \(a=3M\).}
\label{fig:bc}
\end{figure}

\begin{figure}[tb]
\includegraphics[width=\columnwidth]{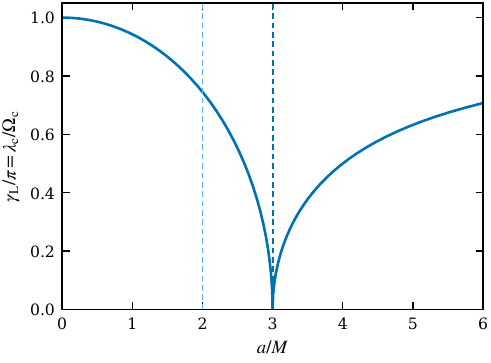}
\caption{Dimensionless instability \(\gamL/\pi=\lam/\Om\). The instability is sensitive to \(a\) throughout the critical-curve-degenerate branch, vanishes at the quartic merger \(a=3M\), and reappears on the supercritical throat branch.}
\label{fig:gamma}
\end{figure}

\begin{table}[tb]
\caption{Analytic null-orbit phase structure of the Simpson--Visser family. For \(a\le2M\), the entry \(R=3M\) refers to the unstable orbit associated with each asymptotic exterior. The throat is listed as an ordinary stationary null orbit only when it lies in a static region.}
\label{tab:phase}
\begin{ruledtabular}
\begin{tabular}{c c c c}
Regime & Horizon & Unstable orbit & Throat \\
\hline
\(0<a<2M\) & yes & \(R=3M\) & nonstatic \\
\(a=2M\) & null & \(R=3M\) & null \\
\(2M<a<3M\) & no & \(x=\pm\sqrt{9M^2-a^2}\) & stable \\
\(a=3M\) & no & \(x=0\) & quartic \\
\(a>3M\) & no & \(x=0\) & unstable
\end{tabular}
\end{ruledtabular}
\end{table}

\section{Near-critical scaling and analytical branch diagnostics}
\label{sec:diagnostics}

The transition at \(a=3M\) has simple critical scaling. On the subcritical side, let
\begin{equation}
\epsilon=3M-a>0.
\end{equation}
Thus \(\epsilon\) is the positive distance in parameter space from the critical value on the subcritical side.
The off-throat photon spheres satisfy exactly
\begin{equation}
x_\gamma^2=\epsilon(6M-\epsilon),
\label{eq:xcriticalscaling}
\end{equation}
so \(|x_\gamma|\propto\epsilon^{1/2}\). The unstable exponent and the stable throat frequency behave as
\begin{align}
\lam&=\frac{\sqrt2}{9M}\sqrt{\frac{\epsilon}{M}}
+\cO\!\left[\frac{1}{M}\left(\frac{\epsilon}{M}\right)^{3/2}\right],
\label{eq:lamscalingminus}\\
\omega_{\rm th}&=\frac{1}{9M}\sqrt{\frac{\epsilon}{M}}
+\cO\!\left[\frac{1}{M}\left(\frac{\epsilon}{M}\right)^{3/2}\right].
\label{eq:omegathscaling}
\end{align}
Thus both characteristic frequencies vanish with a square-root critical exponent \(1/2\), while
\begin{equation}
\bar a\simeq\sqrt{\frac{3M}{2\epsilon}}
\label{eq:abarscalingminus}
\end{equation}
diverges as the marginal point is approached.

On the supercritical side, with \(\epsilon=a-3M>0\),
\begin{align}
\lambda_{\rm th}&=\frac{1}{9M}\sqrt{\frac{\epsilon}{M}}
+\cO\!\left[\frac{1}{M}\left(\frac{\epsilon}{M}\right)^{3/2}\right],
\label{eq:lamscalingplus}\\
\bar a&\simeq\sqrt{\frac{3M}{\epsilon}}.
\label{eq:abarscalingplus}
\end{align}
The logarithmic strong-deflection coefficient therefore diverges from both sides, while at the critical point itself the logarithmic expansion is replaced by the quartic power-law regime. The limit is therefore nonuniform: \(\bar a\to\infty\) as \(a\to3M\) does not represent an infinite logarithmic coefficient at \(a=3M\); rather, the quadratic maximum on which the logarithmic expansion is based degenerates, and the correct critical asymptotics are the quartic power law of Eq.~\eqref{eq:quarticDeflectionScaling}.

A minimal model-dependent inversion follows from the critical-curve degenerate branch. Define
\begin{equation}
\eta\equiv\frac{\lam}{\Om}=\frac{\gamL}{\pi}.
\label{eq:eta}
\end{equation}
The dimensionless quantity \(\eta\) is the instability-to-orbital-frequency ratio; equivalently, \(\gamL=\pi\eta\).
For \(a<3M\), Eq.~\eqref{eq:gammaSVsub} gives
\begin{equation}
\frac{a}{M}=3\sqrt{1-\eta^2}
.
\label{eq:ainversion}
\end{equation}
Thus the critical curve alone fixes no value of \(a\) on this branch, whereas the instability ratio does. If the Simpson--Visser family is assumed, the traversability threshold \(a>2M\) is equivalently
\begin{equation}
\eta<\frac{\sqrt5}{3}
,
\qquad 2M<a<3M.
\label{eq:traversabilitycriterion}
\end{equation}
This criterion is deliberately model conditional: it is not a general horizon test. Its significance is that an exact critical-curve degeneracy within a fixed geometric family can be lifted by a second local null-orbit observable.

\section{Discussion, conclusions, and outlook}
\label{sec:conclusion}

For the areal-radius deformation class studied here, the optical
potential separates the role of the original geometry from that of the
radial deformation. A nondegenerate original photon sphere that survives
away from an extremum of $\Rar(x)$ obeys
\begin{equation}
b_{\rm c}=b_{{\rm c},0},
\qquad
\Omega_{\rm c}=\Omega_0,
\qquad
\lambda_{\rm c}=|\Rar'(x_\gamma)|\lambda_0,
\label{eq:discussionfactorization}
\end{equation}
and its local logarithmic strong-deflection coefficient is
$\bar a=\bar a_0/|\Rar'(x_\gamma)|$. Thus the potential height
and orbital frequency can remain unchanged while the instability
and photon-ring hierarchy retain information about the deformation.
The broader complementarity of these observables is already known;
the result here is their exact transformation within the specified
metric class.

At a reflection-symmetric nondegenerate throat, the radial stability
sign factorizes as $U''(0)=U_{0,R}(R_0)\Rar''(0)$.
When the throat coincides with a extremum of $U_0(R)$, the first nonzero
orders compose multiplicatively: $p=mq$. A marginally unstable
maximum with $p>2$ then gives algebraic critical approach,
$|x|\propto|t|^{-2/(p-2)}$, and reflected strong deflection
$\Delta\phi\propto(b/b_{\rm c}-1)^{-(p-2)/(2p)}$.
The generic collision of a nondegenerate maximum of $U_0(R)$ with a
nondegenerate symmetric throat has $m=q=2$, explaining the quartic
normal form without requiring a separate geodesic calculation for
each deformation.

The Simpson--Visser family provides an explicit application across
its black-hole and wormhole regimes. For $0\leq a<3M$, the
critical impact parameter is $3\sqrt3M$, while
$\lambda_{\rm c}/\Omega_{\rm c}=\sqrt{1-a^2/(9M^2)}$.
A stable throat orbit exists for $2M<a<3M$. At $a=3M$, the
three stationary structures merge into a quartic maximum, with
$|x|\propto|t|^{-1}$ and a deflection divergence proportional to
$(b/b_{\rm c}-1)^{-1/4}$. For $a>3M$, the throat is the unique
unstable circular null orbit and the critical scale depends on $a$.
The vanishing instability and divergent logarithmic coefficient on
approach to $a=3M$ signal a nonuniform limit: the logarithmic
expansion must be replaced, not evaluated at the quartic point.
The parameter inversion from the instability ratio is exact within
this family, but is not a general test for an event horizon.

These statements concern local null-geodesic dynamics with a fixed
asymptotic time normalization. An inherited orbit need not be
accessible from infinity if another potential barrier intervenes;
observed image brightness also requires an emission model. Geodesic
stability supplies no verdict on the kinetic, gradient, or nonlinear
stability of the supporting theory. These distinctions delimit the
physical interpretation of the factorization formulas.

Several extensions follow directly from the results. Charged
black bounces provide an immediate non-Schwarzschild application
\cite{GuoMiao2022}.
The dyonic solutions of Junior et al.~\cite{JuniorLobo2025}
provide matter-supported charged backgrounds in which to test the
geometric relations, after checking membership in the metric class
considered here. If electromagnetic rays are used in a nonlinear
electrodynamics model, their characteristic propagation must also be
checked before identifying their trajectories with null geodesics
of the spacetime metric.
Allowing independent metric functions,
\begin{equation}
\dd s^2=-A(x)\dd t^2+B(x)\dd x^2+C(x)\dd\Omega^2,
\label{eq:generalABC}
\end{equation}
would identify which parts of the inheritance law depend on the
reciprocal radial coefficient and which combinations admit a more
general invariant formulation.
Charged wormhole and black-bounce solutions in hybrid
metric--Palatini gravity, such as those constructed by
R\'ois et al.~\cite{RoisLobo2025}, offer concrete backgrounds for
this wider comparison; the restricted inheritance law cannot be
assumed without checking their metric functions.
Asymmetric and multithroat
geometries would test how degenerate extrema split and which local
branches remain visible from a chosen exterior. Near a merger, a
uniform approximation connecting the logarithmic and power-law
regimes would complement the limiting exponents derived here and
provide a useful basis for global ray tracing.

A further bridge is to wave propagation. For a specified covariant
matter and gravitational action, one can compare the optical
extrema with the actual perturbation potentials, trapped modes,
quasinormal frequencies, and echoes. Such a calculation must retain
the supporting-field perturbations: source ambiguity can change the
wave dynamics without changing the background metric
\cite{YangLan2026,SantosLessaMalufOlmo2026}. It would establish when
the geodesic Lyapunov exponent predicts wave damping and when
additional channels or global barriers invalidate that correspondence.
The present results supply an analytic geometric benchmark for these
extensions while keeping critical-curve inheritance, local
instability, and dynamical viability distinct.

\section*{Acknowledgements}

FSNL and LFDS acknowledge support from the Fundação
para a Ciência e a Tecnologia (FCT) through the research grant UID/04434/2025.

\end{document}